\documentclass[preprint,preprintnumbers,amsmath,amssymb,floatfix,nofootinbib,superscriptaddress,longbibliography]{revtex4-1}
\usepackage[dvips]{graphicx}
\usepackage{url}
\usepackage{subfigure}
\usepackage{xcolor}
\usepackage[colorlinks=true,citecolor=blue,urlcolor=blue,linkcolor=black]{hyperref}
\usepackage[export]{adjustbox}
\usepackage{setspace}
\usepackage{bm, bbold, xcolor, slashed}
\usepackage{amsmath,amsfonts}
\usepackage{braket}
\usepackage{empheq}
\usepackage[compat=1.0.0]{tikz-feynman}
\usepackage{feynmp-auto}
\usepackage{tikz}
\usepackage{slashed}
\usepackage{tensor}
\usepackage{graphicx}
\usepackage{float}

\newcommand{\be}{\begin{equation}}
\newcommand{\ee}{\end{equation}}
\newcommand{\ba}{\begin{array}}
\newcommand{\ea}{\end{array}}

\begin{document} 
\title{The massless limit for massive amplitudes and  the contraction of the little group}
\author{J. Lorenzo Díaz-Cruz}
\email{lorenzdx@gmail.com}
\affiliation{Centro Interdisciplinario de Investigacion y Educacion de la Ciencia, BUAP. Ciudad Universitaria, Puebla, Pue. México.
}
\author{Jonathan Perez Réyes}
\email{jonathan.reyesper@alumno.buap.mx}
\affiliation{Centro Interdisciplinario de Investigacion y Educacion de la Ciencia, BUAP. Ciudad Universitaria, Puebla, Pue. México.
}
\affiliation{Facultad de Ciencias Físico - Matemáticas
Benemérita Universidad Autónoma de Puebla
Apdo. Postal 1364, C.P. 72000, Puebla, Pue. México.}

\author{Jorge Leon Silverio}
\email{its.jorgeleon@gmail.com}
\affiliation{Facultad de Ciencias Físico - Matemáticas
Benemérita Universidad Autónoma de Puebla
Apdo. Postal 1364, C.P. 72000, Puebla, Pue. México.}



\date{\today}


\begin{abstract}
This paper is devoted to study the spin-spinor formalism to deal with amplitudes for massive particles. After presenting the basic formulae and conventions, we evaluate in detail the amplitudes for two specific examples, namely the decay $W \rightarrow l \nu_l$ and the reaction $ e^+ e^- \rightarrow \mu^+ \mu^-$. For each case, we display the amplitudes using charts that represent the flow of spin/helicities from the initial to the final particles, which can be used to calculate the total squared amplitude. One can also study the symmetries of the process by comparing different branches along the charts, which are related by those symmetries. Finally, we study the massless limit of the massive theory, employing the concept of Little Group Contraction (LGC), which was used first by Inonu and Wigner to derive the algebra of the Little Group (LG) for the massless case as the limit of the massive one.
\end{abstract}
\maketitle

\tableofcontents

\section{Introduction}

Results from all high energy physics (HEP) fronts show that quantum field theory (QFT) provides the correct framework to describe relativistic quantum phenomena. Among the infinite number of possible QFTs, one of them is singled out by nature as the correct description of the fundamental constituents and its interactions, namely the the standard model (SM) \cite{Schwartz:2014sze}; these correspond to the quarks and leptons interacting through the strong and electroweak gauge interactions.
The use of hadron colliders has enabled us to probe the SM, including the detection of its last piece, the Higgs boson \cite{Higgs:1964pj, ATLAS:2012yve}. Currently, the Large Hadron Collider (LHC) at CERN is analyzing data to complete the Higgs profile, which so far shows consistency with the SM. The SM tests relay on the calculation of quantum chromodynamics (QCD) effects at higher-orders, which is not a trivial task. Substantial progress has been made in recent years, thanks in part to the development of helicity methods, as well as advanced unitary techniques \cite{Cachazo:2004kj,Bern:1994zx,Dixon:1996wi,Henn:2014yza}.

Besides the practical utility of those calculations for testing the SM, they have provided a deeper understanding of QFT principles.  In the canonical formalism, the concept of a quantum operators and field was thought to be fundamental \cite{Peskin:1995ev,Weinberg:1995mt}, however in the modern approach, the concept of field can be circumvented \cite{Cheung:2017pzi}.
  This approach to QFT is based on the transformations of the quantum states under the Poincare group, which includes Lorentz transformations and space-time translations. As a quantum theory, we can think of states that are eigenstates of the momentum operators.  One could construct realistic quantum field theories  based on the concepts of unitarity, locality, and the transformation properties under Wigner's little group. 

The calculation of amplitudes with Feynman diagrams can be replaced by the use of on-shell methods \cite{Dixon:1996wi, Witten:2003nn, Diaz-Cruz:2015oie}. Such progress has permitted us to obtain results that could be considered perturbative jewels \cite{Elvang:2015rqa}. For instance, the Parke-Taylor formula can describe the amplitude for a maximally helicity violating (MHV) n-gluon process \cite{Parke:1986gb}. Thus, a complete alternative program for QFT has emerged, where Amplitudes are the central object, rather than actions or lagrangians. 
These amplitude methods have focused on the massless case, however massive amplitudes have been considered too, for instance in \cite{Dittmaier:1998nn,Kosower:2004yz,Schwinn:2005pi,Diaz-Cruz:2016ahc}. A further major advance to deal with the massive case has been the extension of those methods, with a new and different understanding \cite{Arkani-Hamed:2017jhn,Ochirov:2018uyq}. The key point is to take advantage of the LG and its representations. 

\begin{figure}[H]
    \centering  
\includegraphics{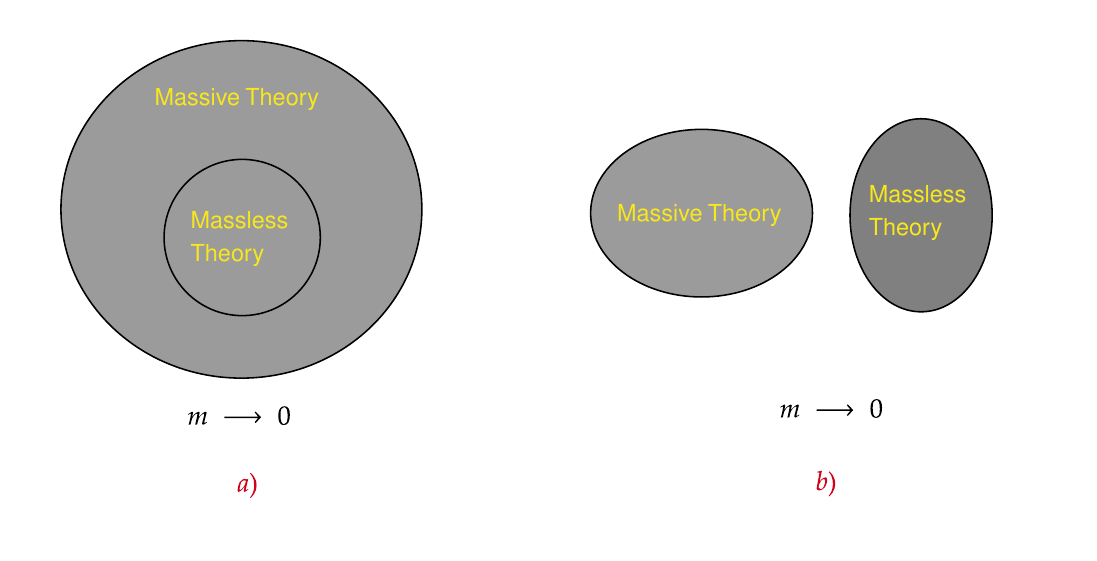}

\caption{Massive-Massless theories}
    \label{Fig1M}
\end{figure}

The Amplitude program includes now  the massive and massless cases,  with methods and concepts that essentially share the origin. Thus, besides discussing how they can be applied to realistic process, 
one can also study how these two cases are connected. Namely, we would like to discuss up to what point a massless QFT can be reachable from the massive one in the limit when some masses vanish, or alternatively in the high-energy limit. This problem was discussed long time ago by Van Dam and Veltman and Zaharov  \cite{vanDam:1970vg,Zakharov:1970cc}, concluding that only in the case of quantum electrodynamics (QED), one can assign a mass to the photon,
and in the limit for a very small photon mass, the theory is the same as the one with a massless photon. This is no longer valid for Yang-Mills theory (YM), even though here the effects appear at the loop-level. Similarly,
for gravitations it was found that the theory with a massive graviton gives different prediction from the massless one, 
thus they are different theories. This is illustrated in Fig. \ref{Fig1M}.
These discussions have been extended to the supergravity case, reaching similar conclusions. Namely, the gravitino amplitudes that include a very small mass are different for the ones with a massless gravitino \cite{Pilling:2004cu,Deser:2000de, Dilkes:2001av}.

In this paper, we would like to discuss these aspects of the massless limit for massive amplitudes from a new perspective; namely, we want to use the concept of Group Contraction as applied to the LG, which permits us to reduce the algebra of $SO(3)$ that describes the spin of massive particles,  to the algebra of the massless case, namely the Euclidean Group $E(2)$. The organization goes as follows: after this introduction (Sect. 1), we present in Sect. 2 the basics of the spin-spinor formalism which describes the quantum states of massive particle of any spin. Then, in Sect. 3 we calculate the amplitudes for two examples, namely  the decay $W\to l \nu_l$ and the reaction $e^+e^- \to \mu^+\mu^-$. We use these examples to discuss some aspects of the corresponding amplitudes, including the description of discrete symmetries and some recipes to calculate the squared amplitude. Then, we discuss in Sect. 4 the concept of Little Group Contraction (LGC) as a mean to understand the massless limit of a massive theory. Finally, conclusions are presented in Sect. 5.


\section{Amplitude methods for massive particles - Spin-Spinors formalism}\label{section2}

In this section, we discuss the work of Arkani-Hammed, Huang and Huang (AH-HH) \cite{Arkani-Hamed:2017jhn}, which makes possible to treat amplitudes for massive particles in a manner that makes transparent the symmetries and quantum numbers of the particles involved.  This was done by extending the helicity spinor formalism suited for massless particles  \cite{Elvang:2015rqa}, which was applied in \cite{DeCausmaecker:1981wzb,Kleiss:1985yh,Xu:1986xb,Diaz-Cruz:2015oie,ReyesPerez:2025akc,Chakraborty:2024sdv}. As one knows, scattering amplitudes transform under the Little group, which is a subgroup of the Lorentz group. Moreover, the little group for massless particles is the helicity group, which is equivalent to a phase transformation, while the little group for massive particles is the rotation  group $SO(3)$ \cite{Wigner:1939cj,Bargmann:1948ck}. The amplitude  transforms as a direct product of the Little group for each external particle appearing in  the amplitude. 

The key to the work of AH-HH was to generalize helicity spinors to a tensor where each row separately transforms under a Lorentz transformation $SL(2,\mathbb{C})$, while each column transforms under the spin group $SU(2)$. That is, they introduce the spin-spinors $\vert \boldsymbol{p}^{I} \rangle$ and $\vert \boldsymbol{p}^{I}]$, which transform under the direct product of spin and Lorentz transformations $SL(2,\mathbb{C})$. For massive particles $\left(p_\mu p^\mu = m^2\right)$, the momentum $p_{\alpha \dot{\alpha}}$ is rank 2, since $\operatorname{det}(p) = m^2$. To describe these momenta, additional indices of $\operatorname{SU}(2)$ are introduced $I = 1,2$, which are contracted with the tensor $\varepsilon_{I J}$. For $I=1$ it corresponds to a particle with spin-z $-\frac{1}{2}$, then $I=2$ it corresponds to a particle with spin-z $\frac{1}{2}$. The massive momentum can then be decomposed as:
\begin{equation}
    p_{\alpha \dot{\alpha}} = \epsilon_{I J} \lambda_\alpha^I \tilde{\lambda}_{\dot{\alpha}}^J 
=  |\boldsymbol{p}^I\rangle_\alpha \left[ \boldsymbol{p}_I \right|_{\dot{\alpha}}, 
\quad 
\bar{p}^{\dot{\alpha} \alpha} = \epsilon^{I J} \tilde{\lambda}_I^{\dot{\alpha}} \lambda_J^\alpha 
=  |\boldsymbol{p}_I]^{\dot{\alpha}} \langle \boldsymbol{p}^I |^{\alpha}.
\end{equation}
Here, $\lambda_\alpha^I$ (angle bracket) corresponds to a left-handed object $\left(\frac{1}{2}, 0\right)$, while $\tilde{\lambda}_{\dot{\alpha}}^J$ (square bracket) corresponds to a right-handed object $\left(0, \frac{1}{2}\right)$. If we separately impose $\operatorname{det}\lambda = \operatorname{det}\tilde{\lambda} = m$, then the action of the little group on $p^\mu$ is given by
\begin{equation}
    \lambda^I_\alpha \mapsto W^{I}_{\;J}\,\lambda^J_\alpha, \quad
\tilde{\lambda}_{I\dot{\alpha}} \mapsto \left(W^{-1}\right)^{\;J}_{I}\,\tilde{\lambda}_{J\dot{\alpha}}.
\end{equation}
In general, $W \in \mathrm{SL}(2,\mathbb{C})$. If $p^\mu$ is real, the little group transformations become $W \in \mathrm{SU}(2)$. For this reason, we refer to $I, J$ as $\mathrm{SU}(2)$ indices, while we refer to $\alpha, \dot{\alpha}$ as Lorentz indices. In this case, the Weyl equations are written as:
\begin{equation}
    p_{\alpha \dot{\alpha}}| \boldsymbol{p}^I \rangle^{\dot{\alpha}}=m | \boldsymbol{p}^I ]_\alpha, \quad \boldsymbol{p}^{\dot{\alpha} \alpha} | \boldsymbol{p}^I]_\alpha=m |\boldsymbol{p}^I \rangle^{\dot{\alpha}},
\end{equation}
For massive particles with mass, $m$, we have $E^2-p^2=m^2$. Now, for a four-vector expressed in spherical coordinates $p^\mu=(E, p \sin \theta \cos \phi, p \sin \theta \sin \phi, p \cos \theta)$, the spin-spinors have the expression:.
\begin{equation}
    |\boldsymbol{p}^I \rangle_\alpha=\left(\begin{array}{cc}
\sqrt{E+p} c & -\sqrt{E-p} s^* \\
\sqrt{E+p} s & \sqrt{E-p} c
\end{array}\right) \quad \text { y } \quad [\left.\boldsymbol{p}_I \right|_{\dot{\alpha} }=\left(\begin{array}{cc}
\sqrt{E+p} c & -\sqrt{E-p} s \\
\sqrt{E+p} s^* & \sqrt{E-p} c
\end{array}\right),
\end{equation}
where $c=\cos \left(\frac{\theta}{2}\right), s=\sin \left(\frac{\theta}{2}\right) \exp (i \phi), s^*=\cos \left(\frac{\theta}{2}\right) \exp (-i \phi)$. Similar to helicity spinors, spin-spinor products such as $\langle\boldsymbol{p}^I \boldsymbol{q}^J\rangle$ or $[\boldsymbol{p}^I \boldsymbol{q}^J]$ are Lorentz invariants but still transform under the spin group $S U(2)$. This is excellent because these spinor products will be useful for constructing scattering amplitudes for massive particles, since they transform similarly under the Little group. Furthermore, mixtures of helicity spinors and spin-spinors such as $\langle p \boldsymbol{q}^J \rangle$ are objects that transform under helicity for the particle $p$ and spin  for the particle $q$. Therefore, with the Weyl helicity spinor formalism and spin spinors, we can write any scattering amplitude of any process, at least in principle.

The reason why the helicity spinors $\vert p \rangle$ and $\vert p ]$ are so useful for the amplitudes of massless particles, is that each of them transforms under a direct product of the helicity group and the Lorentz group $SL(2,\mathbb{C})$. When an inner product is taken, either $\langle pq\rangle$ or $[pq]$, the transformations of $SL(2, \mathbb{C})$ cancel each other exactly, resulting in a Lorentz-invariant product, while the helicity transformations do not. That is, these products transform under a direct product of the helicity transformations for the particles $p_i$ and $p_j$. This is exactly the minimal way needed to write an amplitude involving these particles \cite{Ochirov:2018uyq,Christensen:2018zcq,Christensen:2019mch}.
\begin{figure}[H]
    \centering
\includegraphics{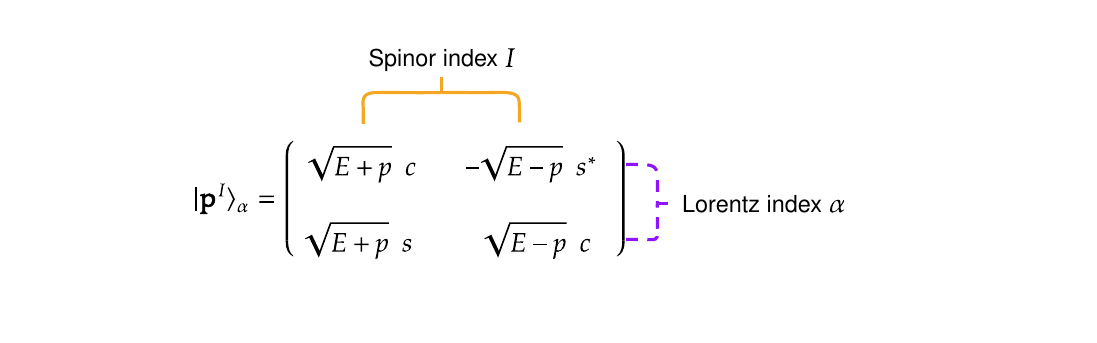}

    \caption{Structure of a spin-spinor. The momentum notation in bold |p⟩, refers to the fact that the momentum of the particle is massive.}
    \label{fig:placeholder}
\end{figure}

We know that spin-spinors are objects that transform under the group $\operatorname{SU}(2)$. It is convenient to define the invariant tensor of this group $\epsilon^{IJ}$, which allows us to define the invariants. For a matrix representation of this group, we say that an object $\psi$, which transforms under $\operatorname{SU}(2)$, transforms as:
\begin{equation}
    \psi^I \rightarrow U^I{ }_J \psi^J, \quad U\in \operatorname{SU}(2),
\end{equation}
where the two fundamental properties that define $\operatorname{SU}(2)$ are:
\begin{equation}
    \epsilon_{I K} U^I{ }_J U^K{ }_L=\epsilon_{J L}, \quad U^I{ }_J\left(U^{\dagger}\right)^J{ }_K=\delta^I{ }_K.
\end{equation}
The tensor $\epsilon_{IJ}$ has the same form as the spinor tensors $\epsilon_{\alpha \beta}$ and $\epsilon_{\dot{\alpha} \dot{\beta}}$, as well as the indices of the group $\operatorname{SU}(2)$. Then, one can rise and lower the indices of  $\psi$ as follows:
\begin{equation}
    \psi_I=\epsilon_{I J} \psi^J \quad \text { and } \quad \bar{\psi}_I=\left(\psi^I\right)^*,
\end{equation}
where the tensors $\epsilon_{\alpha \beta}$, $\epsilon_{\dot{\alpha} \dot{\beta}}$, and $\epsilon_{IJ}$ have the form:
\begin{equation*}
\begin{split}
    \epsilon_{I J}=-\epsilon^{I J}=\epsilon_{\alpha \beta}=-\epsilon^{\alpha \beta}=\epsilon_{\dot{\alpha} \dot{\beta}}=-\epsilon^{\dot{\alpha} \dot{\beta}}=
    \left(\begin{array}{cc}
        0 & -1 \\
        1 & 0
    \end{array}\right).
\end{split}
\end{equation*}
To define all possible combinations of spin-spinors we will generally have to use the tensors $\epsilon_{\alpha \beta}$, $\epsilon_{\dot{\alpha}\dot{\beta}}$, and $\epsilon_{IJ}$ to obtain all possible combinations. If we contract the tensors associated with the Lorentz indices $\langle\boldsymbol{p}^I |^{\alpha }=\epsilon^{\alpha \beta} | \boldsymbol{p}^I \rangle_\beta$ and $| \boldsymbol{p}_I ]^{\dot{\alpha}}=\epsilon^{\dot{\alpha} \dot{\beta}}[\boldsymbol{p}_I |_{\dot{\beta} }$.
\begin{equation}
    \langle\boldsymbol{p}^I |^{\alpha }=
    \begin{pmatrix}
        \begin{array}{cc}\sqrt{E+p} s & \sqrt{E-p} c \\
        -\sqrt{E+p} c & \sqrt{E-p} s^*\end{array}
    \end{pmatrix}, \quad 
    | \boldsymbol{p}_I ]^{\dot{\alpha}}= 
    \begin{pmatrix}
        \begin{array}{cc}
        \sqrt{E+p} s^* & \sqrt{E-p} c \\
        -\sqrt{E+p} c & \sqrt{E-p} s
    \end{array}
    \end{pmatrix}.
\end{equation}
Now we contract and change the position of the indices of $\operatorname{SU}(2)$ by contracting the tensors $\epsilon_{IJ}$. Then $|\boldsymbol{p}_I \rangle_{\alpha}=|\boldsymbol{p}^J \rangle_{\alpha} \epsilon_{\mathrm{JI}}$ and $[\boldsymbol{p}^I |_{\dot{\alpha}}=[\boldsymbol{p}_J |_{\dot{\alpha} } \epsilon^{J I}$.
\begin{equation}
    |\boldsymbol{p}_I \rangle_{\alpha } = 
    \begin{pmatrix}
-\sqrt{E-p} s^* & -\sqrt{E+p} c \\
\sqrt{E-p} c & -\sqrt{E+p} s
    \end{pmatrix}, \quad 
    [\boldsymbol{p}^I |_{\dot{\alpha}} = 
    \begin{pmatrix}
\sqrt{E-p} s & \sqrt{E+p} c \\
-\sqrt{E-p} c & \sqrt{E+p} s^*
    \end{pmatrix},
\end{equation}
The last combination needed is obtained by contracting both tensors  $\epsilon_{\alpha \beta}$, $\epsilon_{\dot{\alpha} \dot{\beta}}$ and $\epsilon_{IJ}$. Then
$\langle\boldsymbol{p}_I |^{\alpha}=\epsilon^{\alpha \beta} | \boldsymbol{p}^J \rangle_\beta \epsilon_{JI}$   and $| \boldsymbol{p}^I ]^{\dot{\alpha}  }=\epsilon^{\dot{\alpha} \dot{\beta}}[\boldsymbol{p}_J |_{\dot{\beta}   } \epsilon^{JI}$. Thereby, we obtain
\begin{align}
  &  \langle\boldsymbol{p}_I |^{\alpha} = \begin{pmatrix}
\sqrt{E-p} c & -\sqrt{E+p} s \\
\sqrt{E-p} s^* & \sqrt{E+p} c
\end{pmatrix},\\ 
& | \boldsymbol{p}^I ]^{\dot{\alpha} }=
\begin{pmatrix}
-\sqrt{E-p} c &\sqrt{E+p} s^* \\
 -\sqrt{E-p} s &-\sqrt{E+p} c
\end{pmatrix}.
\end{align}
In order to perform inner products such as $\langle \boldsymbol{p}^I|^{\alpha } | \boldsymbol{p}^J \rangle_\alpha$, it is necessary to transpose the matrix corresponding to $\langle \boldsymbol{p}^I|^{\alpha}$, so the sum is over the index $I$, explicitly:
\begin{align}
    \begin{split}
        \langle \boldsymbol{p}^I|^{\alpha } | \boldsymbol{p}^J \rangle_\alpha &=\left(\begin{array}{lc}
\sqrt{E+p} s & \sqrt{E-p} c \\
-\sqrt{E+p} c & \sqrt{E-p} s^*
\end{array}\right)^{\operatorname{T}}
\left(\begin{array}{cc}
\sqrt{E+p} c &  -\sqrt{E-p} s^* \\
\sqrt{E+p} s &   \sqrt{E-p} c
\end{array}\right) \\
&=\left(\begin{array}{cc}
0 & -\sqrt{E-p}\sqrt{E+p} \\
\sqrt{E-p}\sqrt{E+p} & 0
\end{array}\right) \\ \nonumber
&= m
\left(\begin{array}{cc}
0 & -1 \\
1 & 0
\end{array}\right)\\ \nonumber
&= -m \epsilon^{IJ}.
    \end{split}
\end{align}
We show some useful products for constructing operators and simplifying amplitudes in the Table \ref{tab:Tabla_Cuadrados_angulados}. These products were calculated explicitly, although they can also be calculated by contracting the $\epsilon$ tensors.
\begin{table}[H]\label{spin spinors table}
\centering
\begin{tabular}{|l|l|l|l|}
\hline
\multicolumn{1}{|c|}{Angle spinors}                                                            & \multicolumn{1}{c|}{square spinors}                                                 & \multicolumn{1}{c|}{angle and square}                                                                                            & \multicolumn{1}{c|}{bi-spinors}                                                                                         \\ \hline
$\langle \mathbf{p}^I \mathbf{p}^J\rangle =-m \epsilon^{\mathrm{I J}}$                       & $[ \mathbf{p}_I \mathbf{p}_J]=-m \epsilon_{ IJ}$ & $|\mathbf{p}^I\rangle_\alpha \langle\left.\mathbf{p}_I\right|^\beta=m \delta_\alpha^\beta$     & $p_{ \alpha \dot{\beta}}=|\mathbf{p}^I\rangle_\alpha [\left.\mathbf{p}_I\right|_{\dot{\beta} }$    \\ \hline
$\langle\mathbf{p}^I \mathbf{p}_J\rangle=-m \delta_{J}^{I}$ & $[\mathbf{p}_I \mathbf{p}^J] =-m \delta_{I}^{J}$ & $|\mathbf{p}_I\rangle_{\alpha } \langle\left.\mathbf{p}^I\right|^{\beta }=-m \delta_\alpha^\beta$                     & $p^{\dot{\alpha} \beta}=| \mathbf{p}_I]^{\dot{\alpha}}\langle\left.\mathbf{p}^I\right|^{\beta}$ \\ \hline
$\langle \mathbf{p}_I \mathbf{p}^J \rangle =m \delta_{I}^{J}$  & $[\mathbf{p}^I  \mathbf{p}_J]=m \delta_{\mathrm{J}}^{\mathrm{I}}$  & $|\mathbf{p}_I]^{\dot{\alpha}} [\left.\mathbf{p}^I\right|_{\dot{\beta}} =m \delta_{\dot{\beta}}^{\dot{\alpha}}$ & $p_{ \alpha \dot{\beta}}=-|\mathbf{p}_I\rangle_{\alpha } [\left.\mathbf{p}^I\right|_{\dot{\beta}} $          \\ \hline
$\langle \mathbf{p}_I \mathbf{p}_J\rangle= m \epsilon_{\mathrm{IJ}}$                        & $[\mathbf{p}^I \mathbf{ p}^J ]=m \epsilon^{\mathrm{IJ}}$                       & $|\mathbf{p}^I]^{\dot{\alpha} } [\left.\mathbf{p}_I\right|_{\dot{\beta}  }= -m \delta_{\dot{\beta}}^{\dot{\alpha}}$              & $p^{\dot{\alpha} \beta}=-| \mathbf{p}^I]^{\dot{\alpha} }\langle\left.\mathbf{p}_I\right|^\beta$    \\ \hline
\end{tabular}
\caption{Internal products of massive spinors.}
\label{tab:Tabla_Cuadrados_angulados}
\end{table}

\bigskip

\subsubsection{Massive spin-$\frac{1}{2}$ particles} In this and the next subsection, we will present the expressions for the Dirac spinors for spin-1/2 particles and the polarization vectors for spin-1 particles. 
For massive particles $\left(p_\mu p^\mu = m^2\right)$ with spin $\frac{1}{2}$, we will find the Dirac equation. For this, it is necessary to construct the operators $ \! \not p + m$ and $\! \not p - m$. First, we adopt the spinor convention
\begin{equation}
|-p\rangle=-|p\rangle, \quad \mid-p]= \mid p], 
\end{equation}
Then one finds the following expression in terms of spin-spinors is
\begin{align}
\gamma_{\mu} p^{\mu}+m=\binom{\left.\mid- \mathbf{p}_{I}\right]^{\dot{\alpha}}}{\left|- \mathbf{p}_{I}\right\rangle_{\alpha}} \otimes \left( \begin{array}{cc}
{[ \mathbf{p}^{I}|_{\dot{\beta}}} & \langle \mathbf{p}^{I}|^{\beta} 
\end{array}\right)=v(-p) \otimes \bar{u}(p).
\end{align}
Thereby, the numerator of the fermionic propagator is reconstructed as follows:
\begin{align}
 \gamma_{\mu} p^{\mu}+m&= \binom{\mid \mathbf{p}_{I}]^{\dot{\alpha}}}{-\left|\mathbf{p}_{I}\right\rangle_{\alpha}} \otimes\left(\begin{array}{cc}
[\mathbf{p}^{I}|_{\dot{\beta}} & \langle \mathbf{p}^{I}|^{\beta}
\end{array}\right) \nonumber\\
&=\left(\begin{array}{cc}
\mid \mathbf{p}_{I}]^{\dot{\alpha}}[ \mathbf{p}^{I}|_{\dot{\beta}} & \mid \mathbf{p}_{I}]^{\dot{\alpha}}\langle \mathbf{p}^{I}|^{\beta}\\
-\left|\mathbf{p}_{I}\right\rangle_{\alpha} [\mathbf{p}^{I}|_{\dot{\beta}}  & -\left|\mathbf{p}_{I}\right\rangle_{\alpha} \langle \mathbf{p}^{I}|^{\beta}
\end{array}\right) \\
&= \left(\begin{array}{cc}
m \delta_{\dot{\beta}}^{\dot{\alpha}} & p^{\dot{\alpha} \beta}  \\
p_{\alpha \dot{\beta}} & m \delta_{\alpha}^{\beta}
\end{array}\right)      .
\end{align}
Therefore, the operators in matrix form are
\begin{equation}
    \not p+m=\left(\begin{array}{cc}
m \delta_\alpha^\beta & p_{\alpha \dot{\beta}} \\
p^{\alpha \beta} & m \delta_{\dot{\beta}}^\alpha
\end{array}\right), \quad \not p-m=\left(\begin{array}{cc}
-m \delta_\alpha^\beta & p_{\alpha \dot{\beta}} \\
p^{\alpha \beta} & -m \delta_{\dot{\beta}}^\alpha
\end{array}\right).
\end{equation}
The Dirac spinors for an outgoing particle and  anti-particle are,
\begin{equation}\label{DiracSS1}
    \bar{u}^I(p)= 
        \begin{pmatrix}
            \langle \boldsymbol{p}^I |^{\alpha } & [ \boldsymbol{p}^I |_{\dot{\alpha}}
        \end{pmatrix}, \quad 
        v_I(p) = 
        \begin{pmatrix}
            | \boldsymbol{p}_I \rangle_{\alpha  } \\
            | \boldsymbol{p}_I]^{\dot{\alpha}}
        \end{pmatrix},
\end{equation}
then we also obtain,
\begin{equation}\label{DiracSS2}
   u_{I}(p)=
        \begin{pmatrix}
            -| \boldsymbol{p}_I \rangle_{\alpha } \\
            | \boldsymbol{p}_I]^{\dot{\alpha}}
        \end{pmatrix}, \quad 
        \bar{v}^I(p)=
        \begin{pmatrix}
            \langle \boldsymbol{p}^I|^{\alpha } & - [ \boldsymbol{p}^I |_{\dot{\alpha}}
        \end{pmatrix}.    
\end{equation}
These spinors fulfill the family identities of the Dirac equation, 
    \begin{equation}
        \begin{split}
        (\not p-m) u^I(p)=0 \quad \text{and} & \quad \bar{u}_I(p)(\not p-m)=0, \qquad \text{for particles}.\\
(\not p+m) v^I(p)=0  \quad \text{and} & \quad \bar{v}_I(p)(\not p+m)=0, \qquad \text{for anti-particles.} \\
        \end{split}
    \end{equation}
and completeness relations,    
\begin{align}
     &\sum_{I=1,2} u^I(p) \bar{u}_I(p)=\not p+m, \\
     &\sum_{I=1,2} v^I(p) \bar{v}_I(p)=\not p-m.  
\end{align}
The quantum nature of these spinors is manifested in the values of the index $I$. If $I=1$, it corresponds to a particle with total spin $-\frac{1}{2}$; for $I=2$, it corresponds to a particle with total spin $\frac{1}{2}$. This is explicitly manifested in the spinor $u_{I}(p)$:
\begin{equation}
    \begin{split}
    &u_{I}(p)=\begin{pmatrix}
        \begin{pmatrix}
            \textcolor{blue}{\sqrt{E+p} s^*} & \textcolor{red}{\sqrt{E-p} c} \\
            \textcolor{blue}{-\sqrt{E+p} c} & \textcolor{red}{\sqrt{E-p} s}
        \end{pmatrix}^{\dot{\alpha}}_I\\
        \begin{pmatrix}
            \textcolor{blue}{\sqrt{E-p} s^*} & \textcolor{red}{\sqrt{E+p} c} \\
            \textcolor{blue}{-\sqrt{E-p} c} & \textcolor{red}{\sqrt{E+p} s}
        \end{pmatrix}_{\alpha I}
            \end{pmatrix}\\
        & \quad \quad \quad \quad\quad\quad\:\: I=\textcolor{blue}{1} \quad \quad \quad I=\textcolor{red}{2}
    \end{split}.
\end{equation}
By slightly improvising with the notation, if we assign specific values to the index $I$, the spinor $u_{I=1}(p) = - u(p)^{-\frac{1}{2}}$, and $u_{I=2}(p) = + u(p)^{+\frac{1}{2}}$. The Dirac spinor with indices of $\operatorname{SU}(2)$ contains the quantum information of the particle. By convention, following ref.\cite{Christensen:2018zcq}, we take all spinors to be of the form $u^{I}= \epsilon^{IJ} u_{J} $ and $\bar{u}^{I}$, so that the amplitudes calculated using Feynman rules have indices in the upper part.

\subsubsection{Massive spin-1 particles}
A massive spin-1 particle has three degrees of polarization for the obvious reason that, in its rest frame of reference, the spin vector can point in three different directions. In a spherical coordinate system, we choose our polarization vectors as \cite{Gherghetta:2024tob}:
\begin{align}
\varepsilon^{\mu +} &=-\frac{1}{\sqrt{2}}(0, \cos\theta \cos\phi - i \sin\phi,\; \cos\theta \sin\phi + i \cos\phi,\; -\sin\theta), \\
\varepsilon^{\mu -} &=\frac{1}{\sqrt{2}}(0, \cos\theta \cos\phi + i \sin\phi,\; \cos\theta \sin\phi - i \cos\phi,\; -\sin\theta), \\
\varepsilon^{\mu 0} &=\frac{1}{m}(p,\; E \sin\theta \cos\phi,\; E \sin\theta \sin\phi,\; E \cos\theta),
\end{align}
that satisfy the conditions
\begin{equation}
    p_\mu \varepsilon^{\mu, \pm, 0} = 0, \qquad 
\varepsilon^{\mu +}\varepsilon_\mu^{-} = 1, \qquad
\varepsilon^{\mu 0}\varepsilon_\mu^{0} = -1, \qquad
\varepsilon^{\mu \pm}\varepsilon_\mu^{\pm} = 0.
\end{equation}
The polarization vectors can be parameterized using indices of $\operatorname{SU}(2)$:
\begin{align}
&\varepsilon_\mu^{IJ} = \frac{1}{\sqrt{2}\, m} \langle \boldsymbol{p}^I | \sigma_\mu | \boldsymbol{p}^J ], 
\qquad
\varepsilon^{IJ}_{\alpha\dot{\alpha}} = \frac{\sqrt{2}}{m} 
|\boldsymbol{p}^I\rangle_{\alpha}\, [\boldsymbol{p}^J|_{\dot{\alpha}}, \\
&\varepsilon_{\alpha \dot{\alpha}}^{-} \equiv \varepsilon_{\alpha \dot{\alpha}}^{11}, \qquad
\varepsilon_{\alpha \dot{\alpha}}^{+} \equiv \varepsilon_{\alpha \dot{\alpha}}^{22}, \qquad
\varepsilon_{\alpha \dot{\alpha}}^{0} \equiv \frac{1}{2}
\left( \varepsilon_{\alpha \dot{\alpha}}^{12} + \varepsilon_{\alpha \dot{\alpha}}^{21} \right),
\end{align}
where the label $p$ corresponds to the four-momentum of the particle. Being a spin-1 particle, it then needs two $SU(2)$ indices.

\section{Massive amplitudes with charts for spin configurations}\label{section3}
We shall discuss now the explicit expressions of massive amplitudes for two examples, namely the decay $W \to l \bar{\nu}$  and the scattering process $e^{+} e^{-} \rightarrow \mu^{+} \mu^{-}$.

\subsection{Decay $W \to l \bar{\nu}$ }
We take into account the quantum states of the decay $W \to l \bar{\nu}$. First, we consider that there are three possible spin-z states $(-1,0,+1)$ for the $W$ boson, two spin-z  states $(-\frac{1}{2}, +\frac{1}{2})$ for the lepton, and only one helicity $(+\frac{1}{2})$ for the antineutrino. Considering that the amplitude $\mathcal{M}(\boldsymbol{1}^{1}, \boldsymbol{2}^{\frac{1}{2}}, 3^{+\frac{1}{2}})$ of this process contains six possible quantum configurations, it can be expressed by means of a spin-helicity configuration chart.

 In these charts the black circle represents the particle (and its momentum), then one attaches a number of lines of different color to represent the possible spin-z components of this particle. At the end of each line one attaches another black circle for the next particle participating in the reaction, and this is repeated till all the particles and spins are considered.

For the decay $W \to l \bar{\nu}$ the resulting chart is the following,
\begin{figure}[H]
    \centering
 \scalebox{0.85}{
\includegraphics{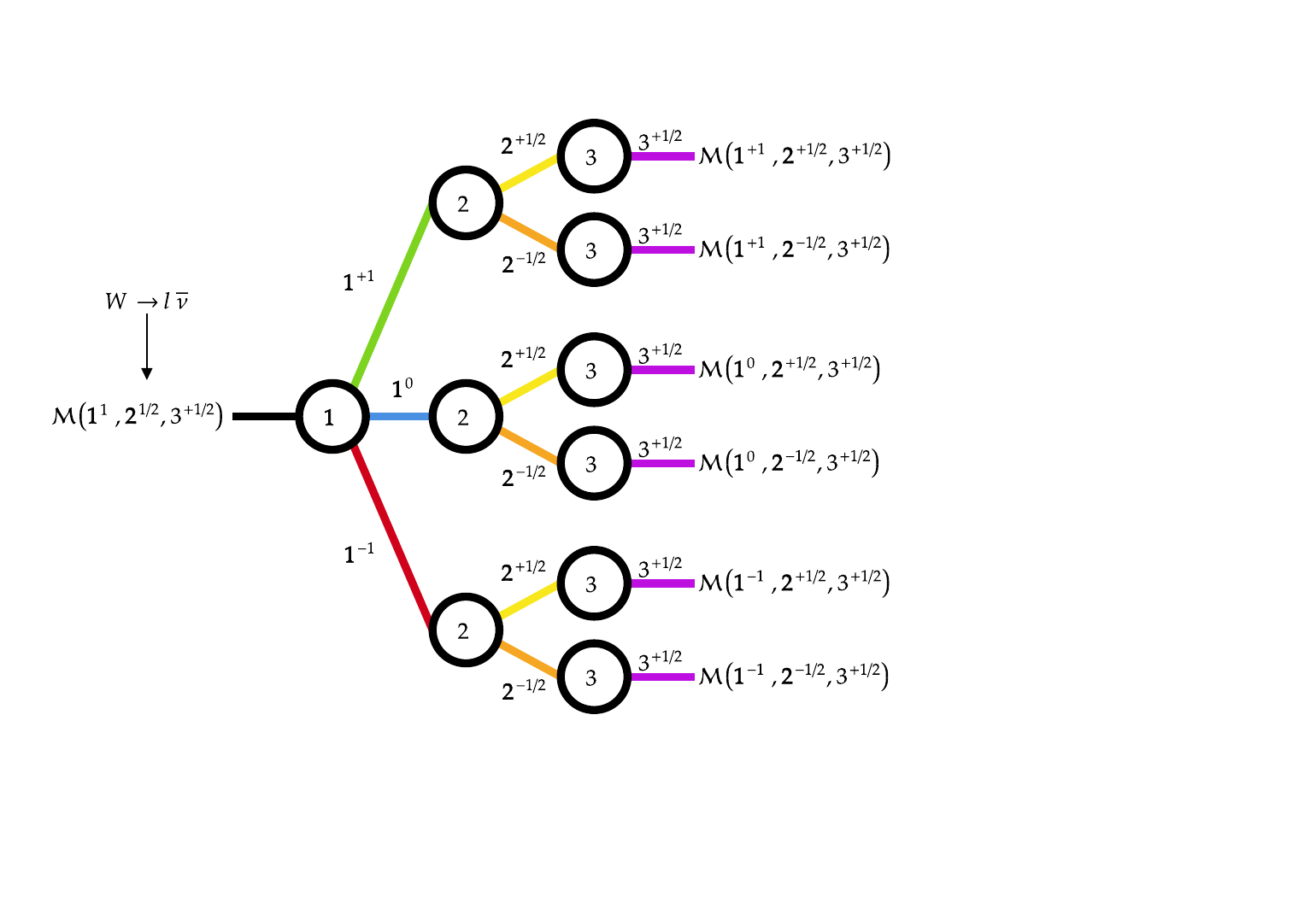}  
}
    \caption{This chart shows the different spin and helicity configurations for the decay $W \to l \bar{\nu}$. The $W$ boson has three  spin-z components (-1, 0, +1), the lepton has two spin-z states (-1/2, +1/2), while the antineutrino has only one helicity state (+1/2). }
    \label{fig:DecayWln}
\end{figure}

The three-point amplitude for this process using on-shell variables is:
\begin{equation}
    \mathcal{M}(\boldsymbol{1}^{1}, \boldsymbol{2}^{\frac{1}{2}}, 3^{+\frac{1}{2}})=\mathcal{M}^{IJK}=
    \frac{g_{W l \nu}}{M_W} \langle \boldsymbol{1}^{I} \boldsymbol{2}^{K} \rangle [3 \boldsymbol{1}^{J}],
\end{equation}
where particle $\boldsymbol{1}$ is the $W$ boson and is represented by two bold spinors whose spin indices are (implicitly) sintered; particle $\boldsymbol{2}$ is the charged lepton, represented by a bold spinor, and particle 3 is the antineutrino, represented by a helicity spinor. We can give explicit values to the indices $I$, $J$, and $K$ to obtain the explicit forms of the amplitude as shown in Table \ref{fig:DecayWln}.

\begin{table}[H]
\begin{tabular}{|c|c|c|c|}
\hline
\begin{tabular}[c]{@{}c@{}}Index \\ configuration$IJK$\end{tabular} & Amplitude  $\mathcal{M}^{IJK} $                                   & \begin{tabular}[c]{@{}c@{}}Square amplitude\\ $|\mathcal{M}^{IJK}|^2$\end{tabular} & \begin{tabular}[c]{@{}c@{}}Spin configuration\\ and associated helicity\end{tabular} \\ \hline
$\mathcal{M}^{111}$                                                       & $A \sin ^2(\theta / 2) \exp (i \phi)$         & $A^2  \sin ^4 (\theta / 2) $                                                   & $\mathbf{1}^{-1}, \mathbf{2}^{-\frac{1}{2}}, 3^{+\frac{1}{2}}$                \\ \hline
$\frac{1}{\sqrt{2}}(\mathcal{M}^{121}+\mathcal{M}^{211})$                 & $- \frac{1}{\sqrt{2}}A \sin \theta $                   & $\frac{1}{2} A^2 \sin^2 \theta $                                                         & $\mathbf{1}^{0}, \mathbf{2}^{-\frac{1}{2}}, 3^{+\frac{1}{2}}$                 \\ \hline
$\mathcal{M}^{221}$                                                       & $A \cos ^2(\theta / 2) \exp (-i \phi)$        & $A^2 \cos ^4(\theta / 2) $                                                     & $\mathbf{1}^{+1}, \mathbf{2}^{-\frac{1}{2}}, 3^{+\frac{1}{2}}$                \\ \hline
$\mathcal{M}^{112}$                                                       & $A B \sin \theta$                             & $A^2 B^2 \sin^2 \theta$                                                        & $\mathbf{1}^{-1}, \mathbf{2}^{+\frac{1}{2}}, 3^{+\frac{1}{2}}$                \\ \hline
$\frac{1}{\sqrt{2}}(\mathcal{M}^{122}+\mathcal{M}^{212})$                 & $- \frac{1}{\sqrt{2}} A B \cos \theta \exp (-i \phi) $ & $ \frac{1}{2} A^2 B^2 \cos^2 \theta  $                                                   & $\mathbf{1}^{0}, \mathbf{2}^{+\frac{1}{2}}, 3^{+\frac{1}{2}}$                 \\ \hline
$\mathcal{M}^{222}$                                                       & $-A B \sin \theta \exp (-2 i \phi)$           & $A^2 B^2 \sin^2 \theta $                                                       & $\mathbf{1}^{+1}, \mathbf{2}^{+\frac{1}{2}}, 3^{+\frac{1}{2}}$                \\ \hline
\end{tabular}
\caption{In this table we show all the possible quantum configurations of the amplitude $W \to l \bar{\nu}$. The coefficients are $A=g_{W l \nu} M_W \sqrt{1 - B^2}$ and $B = \frac{M_l}{M_W}$}
\label{fig:DecayWln}
\end{table}. 
The amplitudes in this explicit form have many advantages. One is the ability to perform CPT transformations by only changing the energy, momentum, and angles. Another is the ability to directly take the high-energy limit of any particle ($p_i \to E_i$), etc. As an example, we can see that if the mass of the W boson is very large compared to that of the lepton, the coefficient $B \to 0$ will eliminate the configurations $1^{-1,0,+1},2^{ +\frac{1}{2}},3^{ +\frac{1}{2}}$, allowing the electron mass to be discarded. Next, we take the square of the absolute value of each term and sum them to obtain the squared Feynman amplitude:
\begin{equation}
    \begin{split}
        \frac{1}{3} \sum_{\text{All spins}} |\mathcal{M}|^2 &= \frac{1}{6} \Bigl( |\mathcal{M}(\mathbf{1}^{-1}, \mathbf{2}^{-\frac{1}{2}}, 3^{+\frac{1}{2}})|^2+|\mathcal{M}(\mathbf{1}^{0}, \mathbf{2}^{-\frac{1}{2}}, 3^{+\frac{1}{2}})|^2+|\mathcal{M}(\mathbf{1}^{+1}, \mathbf{2}^{-\frac{1}{2}}, 3^{+\frac{1}{2}})|^2\\
        &+|\mathcal{M}(\mathbf{1}^{-1}, \mathbf{2}^{+\frac{1}{2}}, 3^{+\frac{1}{2}})|^2+|\mathcal{M}(\mathbf{1}^{0}, \mathbf{2}^{+\frac{1}{2}}, 3^{+\frac{1}{2}})|^2+|\mathcal{M}(\mathbf{1}^{+1}, \mathbf{2}^{+\frac{1}{2}}, 3^{+\frac{1}{2}})|^2 \Bigr)\\
        & = \frac{g_{W l \nu}^2}{6 M_W^2}\left(2 M_W^2+m_l^2\right)\left(M_W^2-m_l^2\right) .
    \end{split}
\end{equation}
A better understanding of the importance of the dynamics of this decay is achieved by studying spin dependence. Although in terms of Feynman diagrams this information is implicit in the total sum of all polarizations, using this method one can use the explcit expression for each term:
\begin{equation}
    \begin{split}
        \frac{d \Gamma}{d \Omega} & = \frac{1}{3} \biggl( \frac{d \Gamma (\mathbf{1}^{-1}, \mathbf{2}^{-\frac{1}{2}}, 3^{+\frac{1}{2}})}{d \Omega} +\frac{d \Gamma (\mathbf{1}^{0}, \mathbf{2}^{-\frac{1}{2}}, 3^{+\frac{1}{2}})}{d \Omega} + \frac{d \Gamma (\mathbf{1}^{+1}, \mathbf{2}^{-\frac{1}{2}}, 3^{+\frac{1}{2}})}{d \Omega} \\
        & \quad \quad \quad + \frac{d \Gamma (\mathbf{1}^{-1}, \mathbf{2}^{+\frac{1}{2}}, 3^{+\frac{1}{2}})}{d \Omega} +\frac{d \Gamma (\mathbf{1}^{0}, \mathbf{2}^{+\frac{1}{2}}, 3^{+\frac{1}{2}})}{d \Omega} +\frac{d \Gamma (\mathbf{1}^{+1}, \mathbf{2}^{+\frac{1}{2}}, 3^{+\frac{1}{2}})}{d \Omega} \biggr)
    \end{split}
\end{equation}

to obtain:
\begin{align}
    \frac{d \Gamma (\mathbf{1}^{-1}, \mathbf{2}^{-\frac{1}{2}}, 3^{+\frac{1}{2}})}{d \Omega}&= \frac{p}{32 \pi^2 M_W^2} \left( A^2 \operatorname{sen}^4 \left(\frac{\theta}{2}\right)  \right)  , \\
    \frac{d \Gamma (\mathbf{1}^{0}, \mathbf{2}^{-\frac{1}{2}}, 3^{+\frac{1}{2}})}{d \Omega}&= \frac{p}{32 \pi^2 M_W^2} \left( \frac{A^2}{2} \operatorname{sen}^2 \theta  \right) ,\\
    \frac{d \Gamma (\mathbf{1}^{+1}, \mathbf{2}^{-\frac{1}{2}}, 3^{+\frac{1}{2}})}{d \Omega}&= \frac{p}{32 \pi^2 M_W^2} \left( A^2 \operatorname{cos}^4 \left(\frac{\theta}{2}\right)  \right) ,\\
    \frac{d \Gamma (\mathbf{1}^{-1}, \mathbf{2}^{+\frac{1}{2}}, 3^{+\frac{1}{2}})}{d \Omega}&=\frac{p}{32 \pi^2 M_W^2} \left( A^2 B^2 \operatorname{sen}^2 \theta  \right) ,\\
    \frac{d \Gamma (\mathbf{1}^{0}, \mathbf{2}^{+\frac{1}{2}}, 3^{+\frac{1}{2}})}{d \Omega}&= \frac{p}{32 \pi^2 M_W^2} \left( \frac{A^2 B^2}{2} \operatorname{cos}^2 \theta  \right),\\
    \frac{d \Gamma (\mathbf{1}^{+1}, \mathbf{2}^{+\frac{1}{2}}, 3^{+\frac{1}{2}})}{d \Omega} &=\frac{p}{32 \pi^2 M_W^2} \left( A^2 B^2 \operatorname{sen}^2 \theta  \right) ,
\end{align}


\subsection{Scattering process $e^{+} e^{-} \rightarrow \mu^{+} \mu^{-}$}

We shall use the hybrid method in order to obtain an expression for this amplitude in terms of spinor variables. This method consists of using  first Feynman's rules to write down the amplitude, then changing variables; that is, we go from four-component spinors to spin-spinors whose columns have two components \cite{JRPABAII}. The amplitude obtained from the Feynman diagram for this process is given by:
\begin{align}
    i \mathcal{M}\left(e^{+} e^{-} \rightarrow \mu^{+} \mu^{-}\right)&=  \bar{v} \left(p_2\right)\left(-i e \gamma^{\mu}\right) u(p_1)\left(\frac{-i g_{\mu v}}{q^{2}}\right) \bar{u}(p_3)\left(-i e \gamma^{v}\right) v \left(p_4\right)  \nonumber \\
    &=\frac{e^2}{s} \left(\bar{v}(p_2) \gamma^\mu u(p_1)\right)\left(\bar{u}\left(p_3\right) \gamma_\mu v\left(p_4\right)\right), \quad 
\end{align}
Here the convention we use for momentum conservation is: $p_1 +p_2 + p_3 + p_4 = 0$, then a Mandelstam variable takes the form: $s=(p_1+p_2)^2 = - (p_3 + p_4)^2$ . In this section we shall express the Dirac spinors in terms of the spin-spinors using \eqref{DiracSS1} and \eqref{DiracSS2}, together with the Dirac matrix in the Weyl representation:
\begin{align}\label{RQuiral}
\gamma^{\mu} = 	\left(
\begin{array}{cc}
	0   &  \sigma^{\mu} \\
	\bar{\sigma}^{\mu} &  0
\end{array} \right),   
\gamma^0 = 		\left(
\begin{array}{cc}
	0   &  \mathbf{1} \\
	\mathbf{1} &  0
\end{array} \right),
\gamma^i = 	\left(
\begin{array}{cc}
	0   &  \sigma^i\\
	- \sigma^i &  0
\end{array} \right),
\end{align}
from Clifford's algebra, we have that
\begin{equation}
	\gamma^0 \gamma^i = - \gamma^0 \gamma_i = 
		\left(
	\begin{array}{cc}
		- \sigma^i  &  0  \\
		0  &  \sigma^i
	\end{array} \right).
\end{equation}
For each particle with spin $s$, we will need $2s$ spin indices $SU(2)$; then for our case, we will need $2(1/1) = 1$ spin indices $I$, since electrons and muons are particles of spin $1/2$. For the positron $e^+$, we assign the spin index $I$; for the electron $e^-$, we assign the index $J$; for the muon $\mu^+$, we assign the index $K$; and for the anti-muon $\mu^-$, we have $L$. We proceed to show the  computation of the amplitude in detail:
\begin{align}
    \mathcal{M}(\mathbf{1}^{\frac{1}{2}}, \mathbf{2}^{\frac{1}{2}}, \mathbf{3}^{\frac{1}{2}}, \mathbf{4}^{\frac{1}{2}}) & = \frac{e^2}{s}
    \begin{pmatrix}
            \langle \boldsymbol{2}|^{J} & [ \boldsymbol{2}|^{J}
        \end{pmatrix},\begin{pmatrix}
0 & \bar{\sigma}^\mu \\
\sigma^\mu & 0
\end{pmatrix}
\begin{pmatrix}
            -| \boldsymbol{1} \rangle^I \\
            | \boldsymbol{1}]^{I}
        \end{pmatrix} 
    \begin{pmatrix}
            \langle \boldsymbol{3}|^{K} & - [ \boldsymbol{3} |_{}^{K}
        \end{pmatrix}\begin{pmatrix}
0 & \bar{\sigma}_\mu \\
\sigma_\mu & 0
\end{pmatrix}
\begin{pmatrix}
            | \boldsymbol{4} \rangle^{L} \\
            | \boldsymbol{4}]^L
        \end{pmatrix} \nonumber \\
    &= \frac{e^2}{s} ([\mathbf{2}|\sigma^\mu| \mathbf{1}\rangle+\langle\mathbf{2}| \bar{\sigma}^\mu |\mathbf{1}]) \times ([\mathbf{3}|\sigma_\mu| \mathbf{4}\rangle+\langle\mathbf{3}| \bar{\sigma}_\mu | \mathbf{4}]) \nonumber \\
    &=\frac{e^2}{s} \bigl([\mathbf{2}|\sigma^\mu| \mathbf{1}\rangle [\mathbf{3}|\sigma_\mu| \mathbf{4}\rangle +[\mathbf{2}|\sigma^\mu| \mathbf{1}\rangle \langle\mathbf{3}| \bar{\sigma}_\mu | \mathbf{4}]+  \langle\mathbf{2}| \bar{\sigma}^\mu |\mathbf{1}] [\mathbf{3}|\sigma_\mu| \mathbf{4}\rangle +\langle\mathbf{2}| \bar{\sigma}^\mu |\mathbf{1}] \langle\mathbf{3}| \bar{\sigma}_\mu | \mathbf{4}] \bigr) \nonumber \\
    &=\frac{e^2}{s} \bigl([\mathbf{2}|\sigma^\mu| \mathbf{1}\rangle [\mathbf{3}|\sigma_\mu| \mathbf{4}\rangle +[\mathbf{2}|\sigma^\mu| \mathbf{1}\rangle [\mathbf{4}|\sigma_{\mu}|\mathbf{3}\rangle+ [\mathbf{1}|\sigma^{\mu}|\mathbf{2}\rangle [\mathbf{3}|\sigma_\mu| \mathbf{4}\rangle +[\mathbf{1}|\sigma^{\mu}|\mathbf{2}\rangle [\mathbf{4}|\sigma_{\mu}|\mathbf{3}\rangle  \bigr) \nonumber \\
    & =\frac{e^2}{s} \bigl( 2 [\mathbf{2}\mathbf{3}]\langle\mathbf{1}\mathbf{4}\rangle + 2 [\mathbf{2} \mathbf{4}] \langle\mathbf{1}\mathbf{3}\rangle + 2   [\mathbf{1} \mathbf{3}]\langle\mathbf{2} \mathbf{4}\rangle +2 [\mathbf{1} \mathbf{4}] \langle\mathbf{2} \mathbf{3} \rangle\bigr) \nonumber\\
    &= \frac{2 e^2}{s}\left(\left\langle \mathbf{1}^I \mathbf{3}^K\right\rangle\left[\mathbf{2}^J \mathbf{4}^L\right]+\left\langle \mathbf{1}^I \mathbf{4}^L\right\rangle\left[\mathbf{2}^J \mathbf{3}^K\right]+\left\langle \mathbf{2}^J \mathbf{3}^K\right\rangle\left[\mathbf{1}^I \mathbf{4}^L\right]+\left\langle \mathbf{2}^J \mathbf{4}^L\right\rangle\left[\mathbf{1}^I \mathbf{3}^K\right]\right)
\end{align}
Then the amplitude, including coupling constants, is
\begin{equation}
    \begin{split}
        \mathcal{M}(\boldsymbol{1}^{\frac{1}{2}}, \boldsymbol{2}^{\frac{1}{2}}, \boldsymbol{3}^{\frac{1}{2}}, \boldsymbol{4}^{\frac{1}{2}}) &= 
        \frac{2e^2}{s} \Bigl( 
        \langle \boldsymbol{1}^{I} \boldsymbol{3}^{K} \rangle [ \boldsymbol{2}^{J} \boldsymbol{4}^{L} ] 
        + \langle \boldsymbol{1}^{I} \boldsymbol{4}^{L} \rangle [ \boldsymbol{2}^{J} \boldsymbol{3}^{K} ] 
        + \langle \boldsymbol{2}^{J} \boldsymbol{3}^{K} \rangle [ \boldsymbol{1}^{I} \boldsymbol{4}^{L} ] 
        + \langle \boldsymbol{2}^{J} \boldsymbol{4}^{L} \rangle [ \boldsymbol{1}^{I} \boldsymbol{3}^{K} ]
        \Bigr)\\
         &= \mathcal{M}^{IJKL}.\\
    \end{split}
\end{equation}
To calculate the conjugate amplitude, the following conjugation rule is needed
\begin{align}
    \Bigl(   \langle    \boldsymbol{p}^I \boldsymbol{q}^J\rangle     \Bigr)^* = \Bigl( \langle \boldsymbol{p}^I \vert^{\alpha} \vert \boldsymbol{q}^J \rangle_{\alpha}       \Bigr)^* = [ \boldsymbol{q}_J \vert_{\dot{\alpha}} \vert \boldsymbol{p}_I ]^{\dot{\alpha}} = [\boldsymbol{q}_J \boldsymbol{p}_I]
\end{align}
\begin{align}
     \Bigl(   \langle    \boldsymbol{p}_I \boldsymbol{q}_J\rangle \Bigr)^*  = [\boldsymbol{q}^J  \boldsymbol{q}^I].
\end{align}
Therefore, the conjugate amplitude of $\mathcal{M}$ will be
\begin{align}
        \mathcal{M}^{*}(\boldsymbol{1}^{\frac{1}{2}}, \boldsymbol{2}^{\frac{1}{2}}, \boldsymbol{3}^{\frac{1}{2}}, \boldsymbol{4}^{\frac{1}{2}}) &= 
        \frac{2e^2}{s} \Bigl( 
        \langle \boldsymbol{4}_{L} \boldsymbol{2}_{J} \rangle [ \boldsymbol{3}_{K} \boldsymbol{1}_{I} ] 
        + \langle \boldsymbol{3}_{K} \boldsymbol{2}_{J} \rangle [ \boldsymbol{4}_{L} \boldsymbol{1}_{I} ] 
        + \langle \boldsymbol{4}_{L} \boldsymbol{1}_{I} \rangle [ \boldsymbol{3}_{K} \boldsymbol{2}_{J} ] 
        + \langle \boldsymbol{3}_{K} \boldsymbol{1}_{I} \rangle [ \boldsymbol{4}_{L} \boldsymbol{2}_{J} ]
        \Bigr) \nonumber \\
        &= \mathcal{M}^{*}_{IJKL}.    
\end{align}
The square amplitude is
\begin{align}
    \frac{1}{4} \sum_{\text{spin}} |\mathcal{M}|^2 =& \frac{e^4}{s^2}\Bigl(\left[1^I 3^K\right]\left\langle 2^J 4^L\right\rangle\left[4_L 2_J\right]\left\langle 3_K 1_I\right\rangle+\left[1^I 3^K\right]\left\langle 2^J 4^L\right\rangle\left[3_K 2_J\right]\left\langle 4_L 1_I\right\rangle  \nonumber \\
        &\quad\:\:+ {\left[1^I 3^K\right]\left\langle 2^J 4^L\right\rangle\left[4_L 1_I\right]\left\langle 3_K 2_J\right\rangle+\left[1^I 3^K\right]\left\langle 2^J 4^L\right\rangle\left[3_K 1_I\right]\left\langle 4_L 2_J\right\rangle } \nonumber \\
        &\quad\:\:+ {\left[1^I 4^L\right]\left\langle 2^J 3^K\right\rangle\left[4_L 2_J\right]\left\langle 3_K 1_I\right\rangle+\left[1^I 4^L\right]\left\langle 2^J 3^K\right\rangle\left[3_K 2_J\right]\left\langle 4_L 1_I\right\rangle } \nonumber \\
        &\quad\:\:+ {\left[1^I 4^L\right]\left\langle 2^J 3^K\right\rangle\left[4_L 1_I\right]\left\langle 3_K 2_J\right\rangle+\left[1^I 4^L\right]\left\langle 2^J 3^K\right\rangle\left[3_K 1_I\right]\left\langle 4_L 2_J\right\rangle } \nonumber \\
        &\quad\:\:+ {\left[2^J 3^K\right]\left\langle 1^I 4^L\right\rangle\left[4_L 2_J\right]\left\langle 3_K 1_I\right\rangle+\left[2^J 3^K\right]\left\langle 1^I 4^L\right\rangle\left[3_K 2_J\right]\left\langle 4_L 1_I\right\rangle } \nonumber \\
        &\quad\:\:+ {\left[2^J 3^K\right]\left\langle 1^I 4^L\right\rangle\left[4_L 1_I\right]\left\langle 3_K 2_{J}\right\rangle+\left[2^J 3^K\right]\left\langle 1^I 4^L\right\rangle\left[3_K 1_I\right]\left\langle 4_L 2_J\right\rangle } \nonumber \\
        &\quad\:\:+ {\left[2^J 4^L\right]\left\langle 1^I 3^K\right\rangle\left[4_L 2_J\right]\left\langle 3_K 1_I\right\rangle+\left[2^J 4^L\right]\left\langle 1^I 3^K\right\rangle\left[3_K 2_J\right]\left\langle 4_L 1_I\right\rangle } \nonumber \\
        &\quad\:\:+ {\left[2^J 4^L\right]\left\langle 1^I 3^K\right\rangle\left[4_L 1_I\right]\left\langle 3_K 2_J\right\rangle+\left[2^J 4^L\right]\left\langle 1^I 3^K\right\rangle\left[3_K 1_I\right]\left\langle 4_L 2_J\right\rangle\Bigr) }   
\end{align}
By reordering terms and using the relationships of Table \ref{spin spinors table}, we obtain the following result
\begin{equation*}
    \begin{split}
    \frac{1}{4} \sum_{\text{spin}} |\mathcal{M}|^2      =&\frac{8e^4}{s^2} \bigl( 
        (p_1 \cdot p_3)(p_2 \cdot p_4) + (p_1 \cdot p_4)(p_2 \cdot p_3) 
        + m_{\mu}^2 (p_1 \cdot p_2) + m_{e}^2 (p_3 \cdot p_4) 
        + 2 m_{e}^2 m_{\mu}^2 
        \bigr) \nonumber \\
        =& \frac{4 e^4}{E^4}\left(E^4+|p_1|^2 |p_3|^2 \cos ^2 \theta+E^2\left(m_e^2+m_\mu^2\right)\right)\nonumber \\
        =& \frac{4 e^4}{E^4}\left(E^4+(E^2-m_e^2) (E^2-m_\mu^2) \cos ^2 \theta+E^2\left(m_e^2+m_\mu^2\right)\right)
    \end{split}
\end{equation*}

\begin{figure}[H]
    \centering
\scalebox{0.75}{
\includegraphics{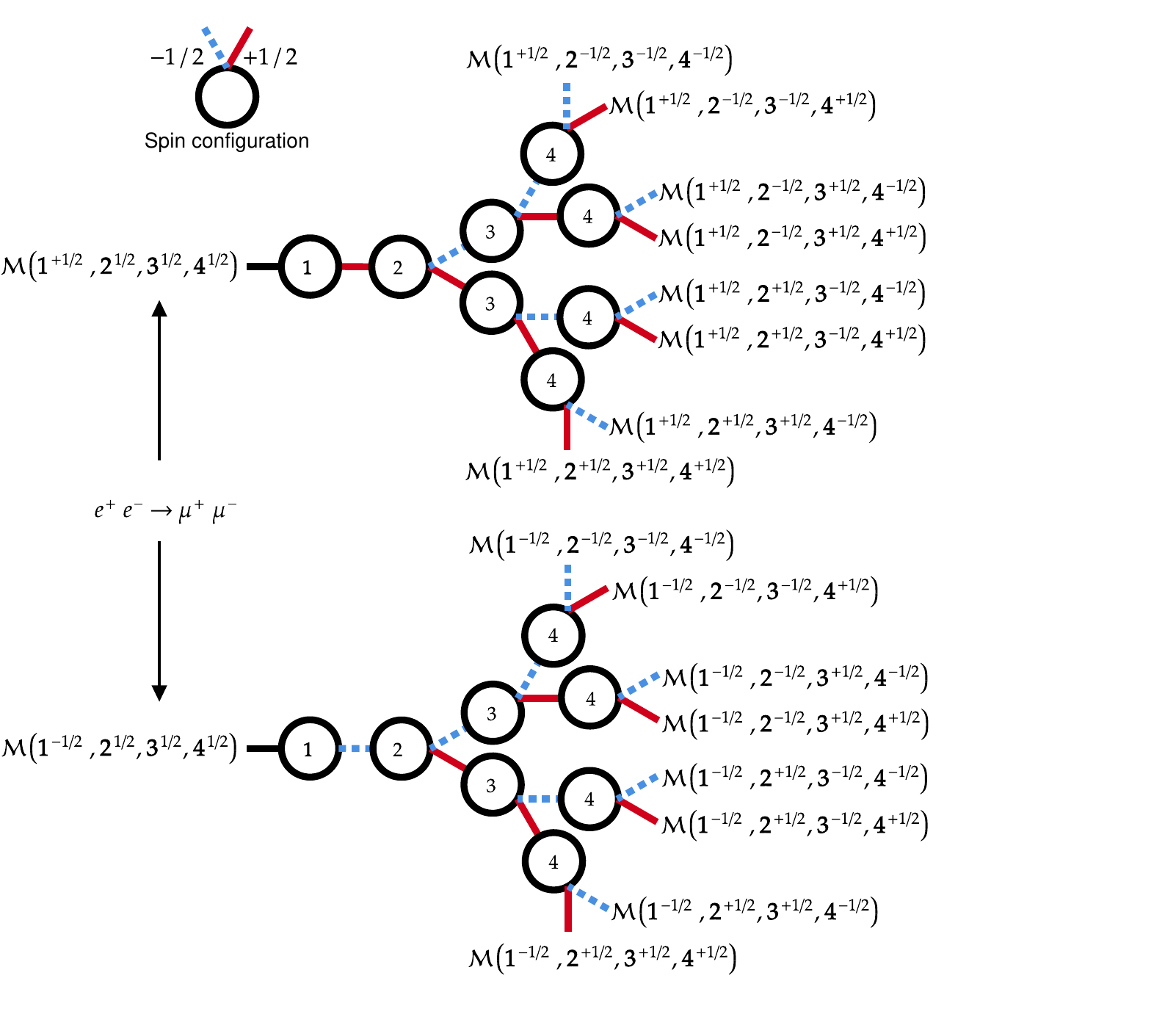}
}
 \caption{All possible spin configuration for $ e^{-} e^{+} \to \mu^{-} \mu^{+} $ starting from particle 1. }
    \label{fig:Anihilationee-mumu}
\end{figure}

\begin{figure}[H]
    \centering  
\scalebox{0.70}{
\includegraphics{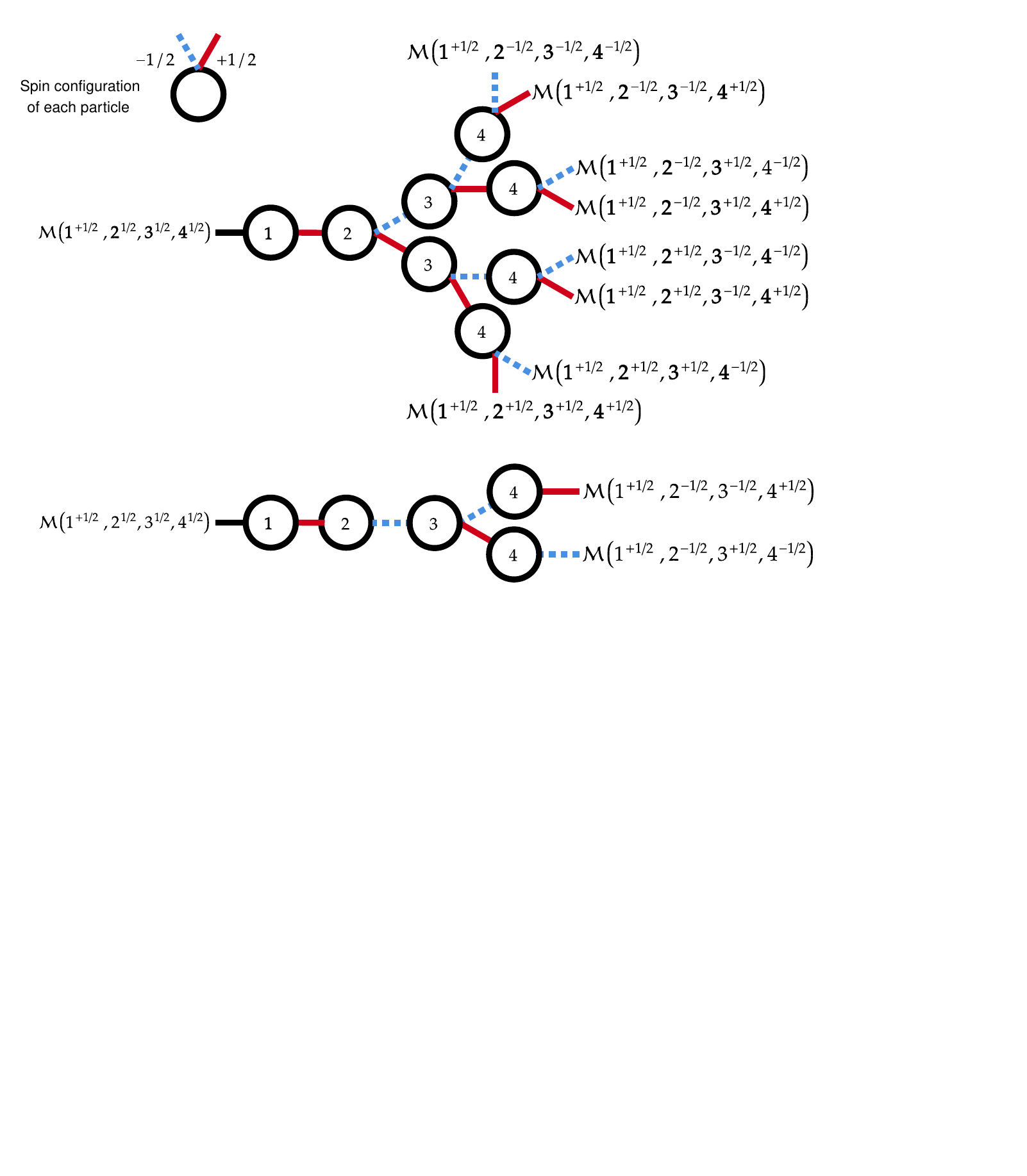}
}

 \caption{All possible spin configuration for $ e^{-} e^{+} \to \mu^{-} \mu^{+} $ starting from particle 2.}
\label{fig:Anihilationee-mumu2}
\end{figure}

On the other hand, following \cite{Christensen:2019mch}, leads us to the following result (in matrix form)  for each spin configuration, :
\begin{align}
    \mathcal{M}(\boldsymbol{1}^{-\frac{1}{2}}, \boldsymbol{2}^{-\frac{1}{2}}, \boldsymbol{3}^{\frac{1}{2}}, \boldsymbol{4}^{\frac{1}{2}}) =& 
    \frac{e^2}{ E^2} 
    \left(\begin{array}{cc}
        m_e m_\mu \operatorname{cos} \theta & -m_e E \operatorname{sin} \theta \\
        -m_e E \operatorname{sin} \theta & -m_e m_\mu \operatorname{cos} \theta
    \end{array}\right), \\[6pt]
    \mathcal{M}(\boldsymbol{1}^{-\frac{1}{2}}, \boldsymbol{2}^{+\frac{1}{2}}, \boldsymbol{3}^{\frac{1}{2}}, \boldsymbol{4}^{\frac{1}{2}}) =& 
    \frac{e^2}{ E^2}
    \left(\begin{array}{cc}
        m_\mu E \sin \theta & -2 E^2 \sin^2\left(\frac{\theta}{2}\right) \\
        2 E^2 \cos^2\left(\frac{\theta}{2}\right) & -m_\mu E \sin \theta
    \end{array}\right) ,\\[6pt]
    \mathcal{M}(\boldsymbol{1}^{+\frac{1}{2}}, \boldsymbol{2}^{-\frac{1}{2}}, \boldsymbol{3}^{\frac{1}{2}}, \boldsymbol{4}^{\frac{1}{2}}) =& 
    \frac{e^2}{ E^2}
    \left(\begin{array}{cc}
        m_\mu E \sin \theta & 2 E^2 \cos^2\left(\frac{\theta}{2}\right) \\
        -2 E^2 \sin^2\left(\frac{\theta}{2}\right) & -m_e E \sin \theta
    \end{array}\right), \\[6pt]
    \mathcal{M}(\boldsymbol{1}^{+\frac{1}{2}}, \boldsymbol{2}^{+\frac{1}{2}}, \boldsymbol{3}^{\frac{1}{2}}, \boldsymbol{4}^{\frac{1}{2}}) =& 
    \frac{e^2}{ E^2}
    \left(\begin{array}{cc}
        -m_e m_\mu \operatorname{cos} \theta & m_e E \operatorname{sin} \theta \\
        m_e E \operatorname{sin} \theta & m_e m_\mu \operatorname{cos} \theta
    \end{array}\right).
\end{align}
If we take $m_e = m_{\mu} =0$ then only $ \mathcal{M}(\boldsymbol{1}^{-\frac{1}{2}}, \boldsymbol{2}^{+\frac{1}{2}}, \boldsymbol{3}^{\frac{1}{2}}, \boldsymbol{4}^{\frac{1}{2}})$ and $ \mathcal{M}(\boldsymbol{1}^{+\frac{1}{2}}, \boldsymbol{2}^{-\frac{1}{2}}, \boldsymbol{3}^{\frac{1}{2}}, \boldsymbol{4}^{\frac{1}{2}})$ survive. All possible spin configurations for this process is shown in chart form in Figs. \ref{fig:Anihilationee-mumu} and
\ref{fig:Anihilationee-mumu2} starting from particles 1 and 2, respectively.

\section{Little group contractions and the massless limit of amplitudes}
\label{section4}
After discussing the  massive amplitudes within the spin-spinor formalism, we would like to discuss the massless limit of the massive QFT. This was done in \cite{Arkani-Hamed:2017jhn} by taking the high-energy limit. Here we shall try to motivate a new perspective, namely we use the concept of Group Contraction, applied to the Little Group. This technique was introduced by Inonu and Wigner \cite{Inonu:1953sp}, and further discussed by Kim and Wigner \cite{Han:1981bn,Kim:1986gq,Kim:1989wt,Kim:2016zus,Baskal:2025waj}, and it permits to obtain the massless Little Group, namely the Euclidean Group $E(2)$, by taking the high-energy limit of the massive Little Group $SO(3)$. We shall argue that a similar limit is behind the High-energy limit for the massive spin-spinors \cite{Arkani-Hamed:2017jhn}. 

To explain the concept of group contraction, we follow refs. \cite{Kim:1989wt, Gilmore:2008zz} where we can see that new Lie algebras can be constructed from old Lie algebras by a process called contraction. This involves re-parameterizing the parameter space of the Lie algebra under consideration, such that the multiplication properties of the group or the commutation relations in the Lie algebra remain well-defined even at some singular limit.
In this contraction process, a new basis set $Y_a$ is related to the original basis set $X_b$ by a change of basis which depends on a parameter $\epsilon$, called squeeze parameter, namely:
\begin{equation}
   Y_{a}= M_{a}{ }^{b}(\epsilon) X_{b}. 
\end{equation}
Some authors call the map M as squeeze transformation \cite{Baskal:2025waj}. Furthermore, structure constants have the transformation properties of a tensor as long as the change-of-basis transformation is non-singular; and then the Lie algebra does not change. The new structure constants are now: 
\begin{equation}
  C_{a b}{ }^{c}(\epsilon)=M_{a}{ }^{i}(\epsilon) M_{b}{ }^{j}(\epsilon)  C_{i j}{ }^{k}\left(M(\epsilon)^{-1}\right)_{k}{ }^{c}. 
\end{equation}
If the transformation is singular, the structure constants $ C_{r s}{ }^{t}(\epsilon)$ should still converge to a well-defined limit
\begin{equation}  
C_{a b}^{d}(0)=\lim _{\epsilon \rightarrow 0} C_{a b}^{d}(\epsilon). 
\end{equation}
It is common for structure constants to exist and to define a Lie algebra that is different from the original. 

Within QFT context, we consider first a massive particle at rest $p^{\mu} = (m,0,0,0)$. In this case the LG is the rotation group $O(3)$, with generators $J_i$. Then we perform a boost along the z axis $B(\eta)$, such that 
\begin{equation}
p^{\mu} \longrightarrow p^{\mu} = (E, 0, 0, E).    
\end{equation}
In this case, $J_3$ remains unchanged, but $J_1$ and $J_2$ are boosted with
\begin{equation}
    J_i \longrightarrow  J'_i = B(\eta) J_i B(- \eta),
\end{equation}
expanding $B(\eta)$ we find
\begin{equation}
   J'_1 = \cosh{\eta} J_1 + \sinh{\eta} K_2, \quad J'_2 =\cosh{\eta}J_2 - \sinh{\eta} K_1. 
\end{equation}
These operators satisfy the same Lie algebra 
\begin{equation}
[J'_i, J'_j ] = i \epsilon_{ijk} J'_k.
\end{equation}
As a result, we define 
 \begin{equation}
     N_1 = - \frac{J´_2}{\cosh{\eta}}, \quad N_2 = \frac{J´_1}{\cosh{\eta}}.
 \end{equation}
So, the massless limit is obtained when the boost parameter $\eta \longrightarrow \infty$; then in this limit, one obtains
\begin{equation}
    N_1 = K_1 - J_2, \quad N_2 = K_2 + J_1,
\end{equation}
we can see that these generators satisfy the Lie algebra
\begin{equation}\label{N1N2Lie}
    [J_3, N_1] = i N_2, \quad [J_3, N_2] = -i N_1, \quad [N_1, N_2] = 0
\end{equation}.
Thus, in the massless limit, this algebra is the same as the $E(2)$ Euclidean group in $2-$dimensions.  On the other hand, we can consider the massive spin-spinors, as written by Ochirov \cite{Ochirov:2018uyq},
\begin{align}
& \lambda_{p \alpha}^{a}= \frac{m}{\sqrt{E+P}}\binom{-e^{-i \varphi} \sin \frac{\theta}{2}}{\cos \frac{\theta}{2}}_{\alpha} \otimes\binom{0}{1}^{a}+\sqrt{E+P}\binom{\cos \frac{\theta}{2}}{e^{i \varphi} \sin \frac{\theta}{2}}_{\alpha} \otimes\binom{1}{0}^{a}, \\
& \tilde{\lambda}_{p}^{\dot{\alpha} a}=\frac{m}{\sqrt{E+P}} \binom{-\cos \frac{\theta}{2}}{-e^{i \varphi} \sin \frac{\theta}{2}}^{\dot{\alpha}} \otimes\binom{1}{0}^{a}+\sqrt{E+P}\binom{e^{-i \varphi} \sin \frac{\theta}{2}}{-\cos \frac{\theta}{2}}^{\dot{\alpha}} \otimes\binom{0}{1}^{a}.
\end{align}
Besides, we want to consider a massive momenta, in the rest frame of the particle it is given by: $p_{\mu}=(m,0,0,0)$, which implies that
\begin{equation}
    P_{\alpha \dot{\alpha}}   = m \left(
	\begin{array}{cc}
		1 &  0 \\
		 0 &  1
	\end{array} \right) = m \left(
	\begin{array}{cc}
		1 &  0 \\
		 0 &  0
	\end{array} \right) + m \left(
	\begin{array}{cc}
		0 &  0 \\
		 0 &  1
	\end{array} \right) .   
\end{equation}
Additionally, this matrix can be written as
\begin{equation}
    P_{\alpha \dot{\alpha}} = \lambda^I_{\alpha} \tilde{\lambda}_{I \dot{\alpha}} = \lambda^1_{\alpha} \tilde{\lambda}_{1 \dot{\alpha}} + \lambda^2_{\alpha} \tilde{\lambda}_{2 \dot{\alpha}},
\end{equation}
here $I=1,2$. Thus, we find that both terms are given by
\begin{equation}
  \lambda^1_{\alpha} \tilde{\lambda}_{1 \dot{\alpha}}  =   m \left(
	\begin{array}{cc}
		1 &  0 \\
		 0 &  0
	\end{array} \right), \quad \lambda^2_{\alpha} \tilde{\lambda}_{2 \dot{\alpha}} = m \left(
	\begin{array}{cc}
		0 &  0 \\
		 0 &  1
	\end{array} \right).
\end{equation}
Therefore, the explicit forms of the spinors are
\begin{equation}
   \lambda^1_{\alpha} = \sqrt{m} \left(
	\begin{array}{c}
		1 \\
		0
	\end{array} \right), \quad \tilde{\lambda}_{1 \ddot{\alpha}} = \sqrt{m} (1,0), \quad \lambda^2_{\alpha} = \sqrt{m}  \left(
	\begin{array}{c}
		0 \\
		1
	\end{array} \right), \quad \lambda_{2 \dot{\alpha}} = \sqrt{m} (0,1).
\end{equation}
Then, performing a similar boost in the z-direction, we obtain  the following for the spinor transformation:
\begin{equation}
    \lambda^{\prime I}_{\alpha} = U \lambda_{\alpha}^I \quad \text{and} \quad \lambda^{\prime}_{I \dot{\alpha}} = \lambda_{I \dot{\alpha}} \tilde{U},
\end{equation}
here, boost matrices $U$ and $\tilde{U}$ have the form
\begin{align}
   & U = \cosh{\frac{\phi}{2}} + \sigma_3 \sinh{\frac{\phi}{2}} =  \left(
	\begin{array}{cc}
		c + s &  0 \\
		 0 &  c -s
	\end{array} \right), \\
 & \tilde{U} = \cosh{\frac{\phi}{2}} - \sigma_3 \sinh{\frac{\phi}{2}} =  \left(
	\begin{array}{cc}
		c - s &  0 \\
		 0 &  c + s
	\end{array} \right),      
\end{align}
where $c= \cosh{\frac{\phi}{2}}$ and $s= \sinh{\frac{\phi}{2}}$. For instance, in the case of $\lambda^I_{\alpha}$ spinors, the boosted spinors are:
\begin{align}
  &\lambda^{ \prime 1}_{\alpha} = U \lambda_{\alpha}^{1}  =   \left(
	\begin{array}{cc}
		c + s &  0 \\
		 0 &  c -s
	\end{array} \right) \times \sqrt{m}  \left(
	\begin{array}{c}
		1 \\
		0
	\end{array} \right) = (c + s) \sqrt{m} \left(
	\begin{array}{c}
		1 \\
		0
	\end{array} \right), \\ 
  & \lambda^{\prime 2}_{\alpha} = U \lambda_{\alpha}^{2}  =   \left(
	\begin{array}{cc}
		c + s &  0 \\
		 0 &  c -s
	\end{array} \right) \times \sqrt{m}  \left(
	\begin{array}{c}
		0 \\
		1
	\end{array} \right) = (c - s) \sqrt{m} \left(
	\begin{array}{c}
		0 \\
		1
	\end{array} \right).
\end{align}
 Furthermore, we know that
 \begin{align}
     \cosh{\frac{\phi}{2}} = \Big( \frac{\gamma +1}{2} \Big)^{1/2}  = \Big( \frac{E +m}{2m} \Big)^{1/2} \quad \text{and} \quad \sinh{\frac{\phi}{2}} = \Big( \frac{\gamma - 1}{2} \Big)^{1/2} =\Big( \frac{E - m}{2m} \Big)^{1/2} ,
 \end{align}
but using Einstein relation $E^2 - p^2 = m^2$ which implies that $(E+m)^{1/2} (E-m)^{1/2} = m$, our relations for the hyperbolic functions become
\begin{align}
    \cosh{\frac{\phi}{2}} = \frac{\sqrt{2}}{(E-m)^{1/2}} \quad \text{and} \quad \sinh{\frac{\phi}{2}} = \frac{\sqrt{2}}{(E +m)^{1/2}},
\end{align}
so, when we take $m=0$, then $\cosh{\frac{\phi}{2}} = \sinh{\frac{\phi}{2}} = \frac{\sqrt{2}}{ \sqrt{E}}$. As a result, we see that $c+s = \frac{2 \sqrt{2}}{\sqrt{E}}$ and $ c -s = 0$, thereby, one kind of spinor survives:
\begin{align}
    \lambda^{\prime 1}_{\alpha} \neq 0 \quad \text{and} \quad \lambda^{\prime 2}_{\alpha} \rightarrow 0
\end{align}
We notice that this is the same form as derived by Ochirov, and similar to the one of AH-HH.


\section{Conclusions}
We have studied the spin-spinor  technique for massive amplitudes, and applied it to calculate HEP processes. In particular, we calculated all the helicity configurations for the decay $W \rightarrow l \nu$ and the reaction $e^+ e^- \rightarrow  \mu^+ \mu^-$. 
In principle, the tools needed to evaluate these processes have been presented in the literature, but these general formulae needs to be expanded in order to get physical information suited for phenomenology, which we hope is the merit of or work, along with other papers presented in the literature \cite{Ochirov:2018uyq,Christensen:2018zcq,Christensen:2019mch}.

In this paper we have represented the amplitudes using charts that describe the flow of spin/helicity from the initial to the final particles.  In these charts we use a black circle to represent the particle (and its momentum), then a number of lines are attached to the circle, using different color to represent the possible spin-z components of this particle. At the end of each line one attaches another black circle for the next particle participating in the reaction, and this is repeated till all the particles and spins are considered. These charts help on a number of details that describe the process, from symmetries to the calculation of the total amplitude.  

We also studied the limit of the massive amplitudes approaching  the massless ones. This has been done in the literature by taking the high energy limit and identifying the components of the spin-spinors that  survive in this limit. Here we point out that this process actually corresponds to the so called Little group contraction (LGC) which we reviewed in section \ref{section4}.

Using LG labels, one can write the massive amplitude for N particles, as follows:
\begin{equation}
    \mathcal{M}^h_{IJ} = \frac{1}{\sqrt{2}m} \lambda_{1 I}^{\alpha} \lambda^{\beta}_{2J} \mathcal{M}^h_{\alpha \beta}
\end{equation}
In principle, when one takes the massless limit for particle K, we just need to apply the LGC and the amplitude should reduce to the massless case \cite{Gomez-Laberge:2025yid}, as we have sown for the processes studied in this paper.

\bigskip


\newpage

\bigskip

{\bf{Acknowledgments.}}
 J.L.D.C would like to thank the support of SECIHTI  and SNI (Mexico). J.R.P thanks for the support of the SECIHTI Scholarship. J.L.S thanks for the support of the SECIHTI  Scholarship .



\bibliography{LG}

\end{document}